\documentclass[sigconf, nonacm]{acmart}

\AtBeginDocument{%
  }

\copyrightyear{2025}
\acmYear{2025}
\setcopyright{rightsretained}
\acmConference[L@S '25]{Proceedings of the Twelfth ACM Conference on Learning @ Scale}{July 21--23, 2025}{Palermo, Italy}
\acmBooktitle{Proceedings of the Twelfth ACM Conference on Learning @ Scale (L@S '25), July 21--23, 2025, Palermo, Italy}\acmDOI{10.1145/3698205.3733948}
\acmISBN{979-8-4007-1291-3/2025/07}

\begin{document}

\title{VTutor for High-Impact Tutoring at Scale: Managing Engagement and Real-Time Multi-Screen Monitoring with P2P Connections}

\author{Eason Chen}
\affiliation{%
  \institution{Carnegie Mellon University}
  \city{Pittsburgh}
  \state{PA}
  \country{USA}
}
\email{eason.tw.chen@gmail.com}

\author{Xinyi Tang}
\affiliation{%
  \institution{Carnegie Mellon University}
  \city{Pittsburgh}
  \state{PA}
  \country{USA}
}

\author{Aprille Xi}
\affiliation{%
  \institution{Carnegie Mellon University}
  \city{Pittsburgh}
  \state{PA}
  \country{USA}
}

\author{Chenyu Lin}
\affiliation{%
  \institution{New York University}
  \city{New York}
  \state{NY}
  \country{USA}
}

\author{Conrad Borchers}
\affiliation{%
  \institution{Carnegie Mellon University}
  \city{Pittsburgh}
  \state{PA}
  \country{USA}
}

\author{Shivang Gupta}
\affiliation{%
  \institution{Carnegie Mellon University}
  \city{Pittsburgh}
  \state{PA}
  \country{USA}
}

\author{Jionghao Lin}
\affiliation{%
  \institution{The University of Hong Kong}
  \city{Hong Kong}
  \country{Hong Kong}
}

\author{Kenneth R Koedinger}
\affiliation{%
  \institution{Carnegie Mellon University}
  \city{Pittsburgh}
  \state{PA}
  \country{USA}
}

\renewcommand{\shortauthors}{Eason Chen, et al.}

\begin{abstract}
Hybrid tutoring, where a human tutor supports multiple students in learning with educational technology, is an increasingly common application to deliver high-impact tutoring at scale. However, past hybrid tutoring applications are limited in guiding tutor attention to students that require support. Specifically, existing conferencing tools, commonly used in hybrid tutoring, do not allow tutors to monitor multiple students' screens while directly communicating and attending to multiple students simultaneously.
To address this issue, this paper introduces \textit{VTutor}, a web-based platform leveraging peer-to-peer screen sharing and virtual avatars to deliver real-time, context-aware tutoring feedback at scale. By integrating a multi-student monitoring dashboard with AI-powered avatar prompts, VTutor empowers a single educator or tutor to rapidly detect off-task or struggling students and intervene proactively, thus enhancing the benefits of one-on-one interactions in classroom contexts with several students. Drawing on insight from the learning sciences and past research on animated pedagogical agents, we demonstrate how stylized avatars can potentially sustain student engagement while accommodating varying infrastructure constraints. Finally, we address open questions on refining large-scale, AI-driven tutoring solutions for improved learner outcomes, and how VTutor could help interpret real-time learner interactions to support remote tutors at scale. The VTutor platform can be accessed at \url{https://ls2025.vtutor.ai}. The system demo video is at \url{https://ls2025.vtutor.ai/video}.
\end{abstract}

\begin{CCSXML}
<ccs2012>
   <concept>
       <concept_id>10003120.10003121.10003129</concept_id>
       <concept_desc>Human-centered computing~Interactive systems and tools</concept_desc>
       <concept_significance>500</concept_significance>
       </concept>
   <concept>
       <concept_id>10010405.10010489.10010493</concept_id>
       <concept_desc>Applied computing~Learning management systems</concept_desc>
       <concept_significance>500</concept_significance>
       </concept>
 </ccs2012>
\end{CCSXML}

\ccsdesc[500]{Human-centered computing~Interactive systems and tools}
\ccsdesc[500]{Applied computing~Learning management systems}

\keywords{High-Impact Tutoring, Tutoring at Scale, Multi-Student Monitoring, Animated Pedagogical Agents, Virtual Avatars}

\maketitle

\section{Introduction}
\label{sec:intro}
High-impact tutoring has emerged as a powerful strategy for boosting student performance and engagement across K--12 and higher education \cite{robinson2021high, white2021early}. Personalized instruction remains a cornerstone of effective learning, yet scaling small-group or one-on-one tutoring to large classes is a persistent challenge \cite{cortes2025scalable, makori2024scaling}. 

Recent research has made progress on scaling one-on-one tutoring in hybrid settings to deliver high-impact tutoring at scale \cite{han2024improving,lin2023personalized,wang2024tutor}. In these contexts, one tutor attends to multiple students through a conferencing tool such as Zoom \cite{lin2023personalized} while students work in educational technologies. These educational technologies and the log data they produce, can generate learning analytics that can help tutors select which student to help (e.g., disengagement and struggle) \cite{beck2013wheel,holstein2022designing,binh2019detecting}. However, these traditional online learning platforms and analytics typically offer limited opportunities to directly communicate with the tutored student, forcing educators to juggle multiple browser windows or find out who needs help by talking to several students in a row. These constraints often prevent timely interventions, which are critical to sustaining motivation and mitigating frustration \cite{wan2019pedagogical,sinclair2003facilitating}. To address these issues, we introduce \textbf{VTutor}, a web-based system that unites:
\begin{enumerate}
    \item \textbf{Real-time multi-student screen sharing} (via WebRTC) so a single tutor can observe and guide many learners concurrently.
    \item \textbf{Interactive avatar tutoring}, leveraging stylized virtual characters to deliver just-in-time feedback and maintain learner engagement.
\end{enumerate}

Drawing on insights from intelligent tutoring systems (ITS) \cite{baker2005designing, vanlehn2011}, teacher augmentation dashboard \cite{holstein2018classroom, holstein2019impact}, and animated pedagogical agents \cite{cassell2000embodied,veletsianos2014pedagogical,woolf2010effect,Lin2024} (as elaborated below), VTutor aims to replicate personalized tutoring's effectiveness at scale. The system is fully browser-based, enabling teachers and tutors to visualize each student’s ongoing work as a grid of thumbnails. Whenever the system or the teacher notices a student is struggling or off-task, they can invoke a stylized avatar on that student’s screen and send them a message. This message is then spoken and displayed on the student's end, featuring messages such as adaptive hints, nudges, or motivational prompts. This approach seeks to reduce the tutor’s cognitive load while maintaining an interactive and interpersonal experience for each learner.

\section{Related Work}
\label{sec:relatedwork}

\begin{figure*}[t]
  \centering
  \includegraphics[width=1\linewidth]{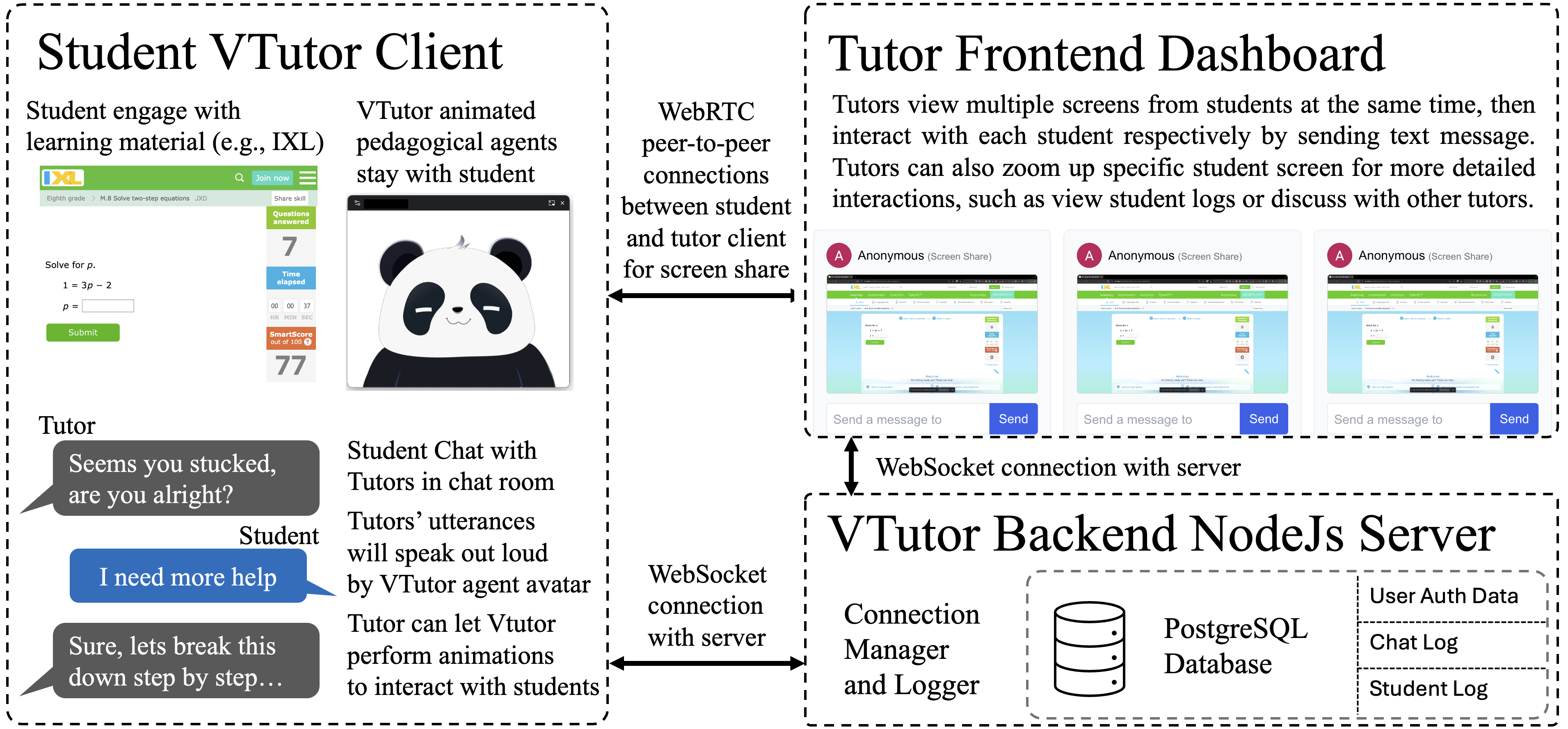}
  \caption{\textbf{VTutor System Architecture.} The \textit{VTutor Student Client} (left) shows how learners interact with educational technologies or learning materials (e.g., IXL) alongside a virtual agent. A chat interface enables conversation with remote tutors, whose utterances can be spoken aloud by the avatar. Tutors use a \textit{Tutor Frontend Dashboard} (top-right) to view multiple student screens simultaneously and send real-time prompts through the avatar. Communication channels include WebSocket connections with backend (for chat, logs, events) and WebRTC peer-to-peer connections (for screen sharing). The \textit{VTutor Backend Node.js Server} (bottom-right) handles user authentication, session management, and logging with a PostgreSQL database.}
  \label{fig:architecture}
\end{figure*}

\subsection{Intelligent Tutoring Systems and Real-Time Teacher Dashboards}
Traditional Intelligent Tutoring Systems (ITS) have demonstrated that promptly detecting student behaviors such as ``gaming the system'' \cite{baker2005designing} or lack of engagement can significantly improve learning outcomes. By identifying off-task patterns and delivering immediate, adaptive feedback, ITS research shows that learners remain more motivated and less likely to persist in unproductive behaviors \cite{baker2005designing, vanlehn2011}. Yet, most existing solutions excel in guiding individual learners rather than scaling oversight to whole classrooms with multiple simultaneous activities. In typical remote teaching settings, teachers and tutors, usually delivered through video conferencing software such as Zoom, are constrained by having to toggle among breakout rooms or multiple shared screens. This constraint reduces the continuous presence of tutors that can effectively curb counterproductive behaviors.

To address these challenges, recent work on teacher dashboards emphasizes the power of immediate, data-informed decisions conveyed to teachers and human tutors \cite{holstein2018classroom, holstein2019impact}. Such dashboards enable educators to identify which learners need attention in real time, prompting more effective interventions. Building on these insights, \textbf{VTutor} consolidates many live student screens within a single interface, drawing on fundamental ITS principles for detecting off-task or ``gaming'' behaviors. When a concerning pattern emerges, VTutor quickly dispatches a targeted nudge or clarification through an animated avatar, preserving the sense of continuous teacher presence. By blending ITS-inspired analytics with a teacher-facing dashboard, VTutor aims to sustain engagement, reduce unproductive shortcuts, and ultimately raise learning outcomes at scale.

\subsection{Pedagogical Agents}
Pedagogical Agents (PAs) leverage on-screen avatars to foster social presence and emotional engagement \cite{cassell2000embodied, porayskapomsta2000providing}. Research shows that well-designed virtual characters can help learners maintain motivation, reduce off-task behavior, and engage in metacognitive reflection \cite{veletsianos2014pedagogical,baker2006adapting}. Similar effects have been observed for one-on-one teacher attention during learning with educational software \cite{karumbaiah2023spatiotemporal}, which our system aims to facilitate. In tutoring contexts, PAs have been used to provide tutoring hints, scaffolding, and emotional support \cite{Domagk2010, Veletsianos2012, Makransky2018,woolf2010effect}.

VTutor employs a stylized avatar with real-time lip-sync and gestures to personalize interactions for each learner, even when a single tutor oversees many students. This approach merges PAs principles with large-scale screen oversight, potentially broadening the impact of virtual tutors in everyday classroom settings.

\section{System Implementation and Architecture}
\label{sec:system-architecture}

The VTutor platform can be access at \url{https://ls2025.vtutor.ai}.
Figure~\ref{fig:architecture} illustrates the three main VTutor components: the \textbf{VTutor Student Client}, the  \textbf{Tutors Frontend Dashboard}, and the \textbf{Node.js Backend Server}.

\noindent\paragraph{VTutor Student Client (Left)}
Students engage with external learning content (e.g., the IXL platform) while a stylized \emph{panda} avatar, referred to as ``VTutor,'' remains visible. This avatar can speak tutor messages aloud, perform animations (e.g., waving), and display textual hints in a speech bubble. The student also has access to a chat box for interacting with the tutor directly.

\noindent\paragraph{Tutors Frontend Dashboard (Top-Right)}
Educators see an array of thumbnails, each representing a student’s screen shared via WebRTC. From this centralized view, tutors can:
\begin{enumerate}
    \item Watch multiple student feeds in low resolution to detect off-task or struggling behavior.
    \item Expand a specific feed to inspect a student’s progress more closely.
    \item Send real-time messages that the avatar speaks or displays for a selected student.
\end{enumerate}
The dashboard additionally provides logs and alerts (e.g., inactivity warnings, repeated incorrect answers) to help tutors pinpoint intervention needs.

\paragraph{Node.js Backend Server (Button-Right)}
A Node.js server manages both user authentication and event logging (e.g., chat logs, student logs). It establishes a persistent WebSocket connection to handle low-latency interactions such as:
\begin{itemize}
    \item Transmission of tutor messages to students’ avatars.
    \item Logging of user events (e.g., ``Student X was inactive for 2 minutes'').
    \item Synchronizing state across the tutor dashboard and all student clients.
\end{itemize}
A PostgreSQL database stores relevant session data (e.g., user credentials, chat transcripts, and student performance logs).

\subsection{WebRTC Multi-Screen Sharing}
\label{sec:webrtc}
VTutor extends the WebRTC framework \cite{petrangeli2019scalable,loreto2014real} to enable multi-student screen sharing with minimal latency. Each student’s local machine encodes an encrypted video feed of their screen, which is then streamed directly to the tutor’s browser (peer-to-peer), with the Node.js server acting as the signaling intermediary. This avoids the computational overhead of a central video-processing server. To optimize performance, VTutor employs adaptive bitrates and selectively requests high-resolution streams only for enlarged (``zoomed-in'') student feeds.

\section{User Flow and Interface}
\label{sec:userflow}

\subsection{Animated Pedagogical Avatar}
\label{sec:avatar}
The ``VTutor'' avatar \cite{chen2025vtutoropensourcesdkgenerative,chen2025vtutoranimatedpedagogicalagent} employs a dedicated animation engine to synchronize lip movements with text-to-speech (TTS) outputs, providing auditory feedback that complements on-screen text. Basic gestures—such as waving, nodding, or a ``thumbs up'' sign—are tied to semantic cues in the tutor’s chat messages (e.g., \emph{encouragement}, \emph{corrective feedback}). By overlaying the avatar on top of learning platforms like IXL, students receive immediate and contextual guidance, maintaining higher levels of engagement.

\begin{figure}[h]
    \centering
    \includegraphics[width=\linewidth]{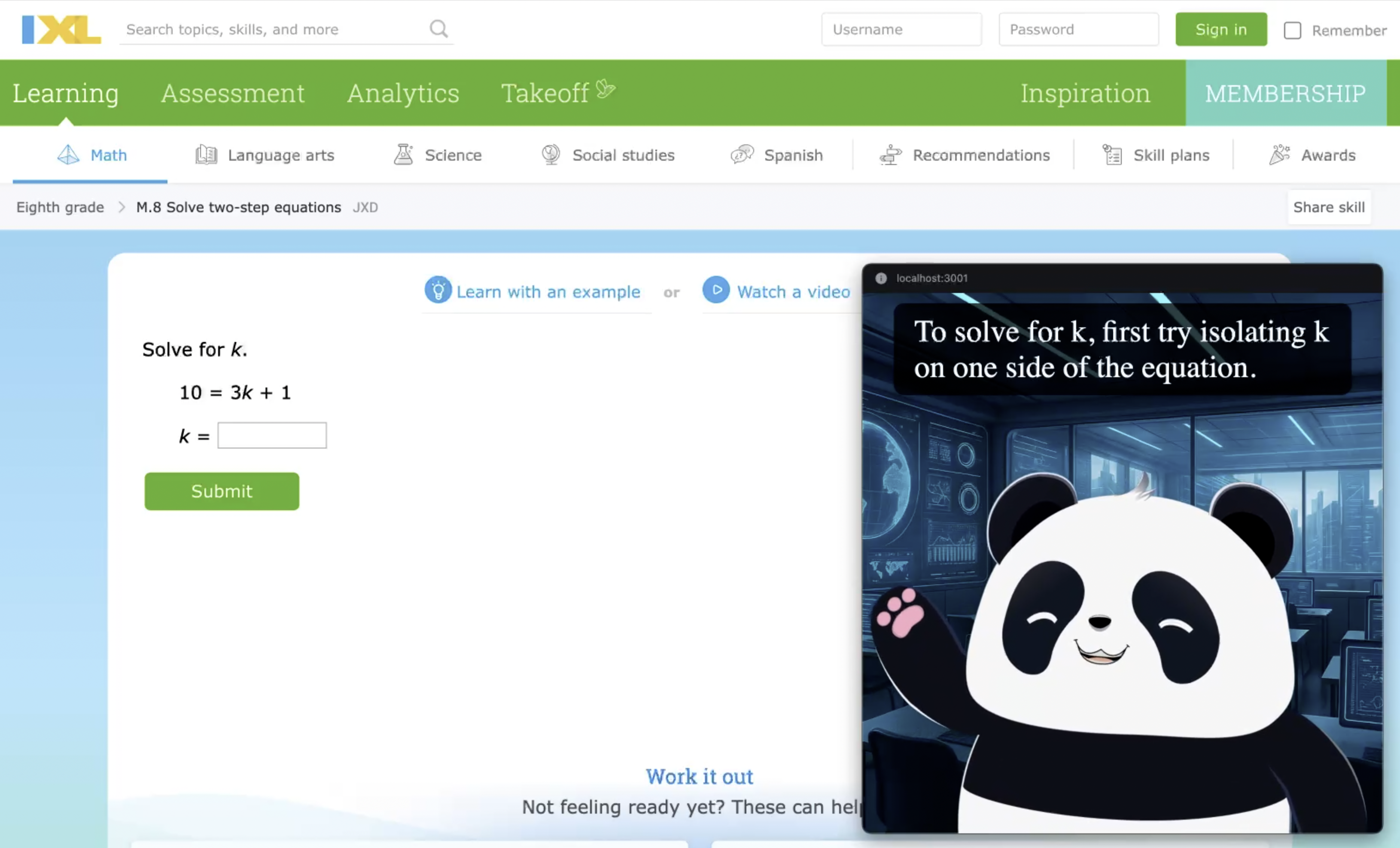}
    \caption{\textbf{Student Interface During Tutoring Session.}
    The student is solving an algebraic equation on IXL (``Solve for $k$''), 
    while the VTutor avatar, an animated panda agent, waves and offers 
    animated guidance in the lower-right corner. Students can also chat 
    directly with tutors; messages from tutors is spoken aloud by the 
    VTutor avatar.}
    \label{fig:student_screenshot}
\end{figure}

Figure~\ref{fig:student_screenshot} shows the student’s perspective during a typical algebra problem. The left region features IXL’s math practice window, where the student enters their solution and clicks ``Submit.'' On the right, the \textit{VTutor} panda avatar is displayed in a pop-up, greeting the learner and prompting them to proceed step-by-step. A chat bubble may appear above the avatar containing the tutor’s text message; simultaneously, the same message is converted into speech (with lip-sync animation).

Below, we describe the typical user flow, referencing both Figures~\ref{fig:architecture} and \ref{fig:student_screenshot} for clarity:

\subsection{Tutor Setup}
The tutor first logs into the VTutor dashboard via a standard web browser and selects an existing class session or creates a new one. The dashboard is divided into a series of student thumbnails, each corresponding to an incoming WebRTC stream. Initially, these thumbnails will remain blank until students join.

\subsection{Student Login and Screen Sharing}
Students access the VTutor platform through a provided URL, enter minimal identifying information (e.g., name or alias), and select the class session. They are prompted by the browser to share their screen or a specific application window. Upon granting permission:
\begin{itemize}
    \item Each student’s screen feed appears as a thumbnail on the tutor’s dashboard.
    \item The avatar window launches locally, ensuring the \textit{VTutor} character is visible and can speak or animate.
\end{itemize}

\begin{figure}[h]
    \centering
    \includegraphics[width=\linewidth]{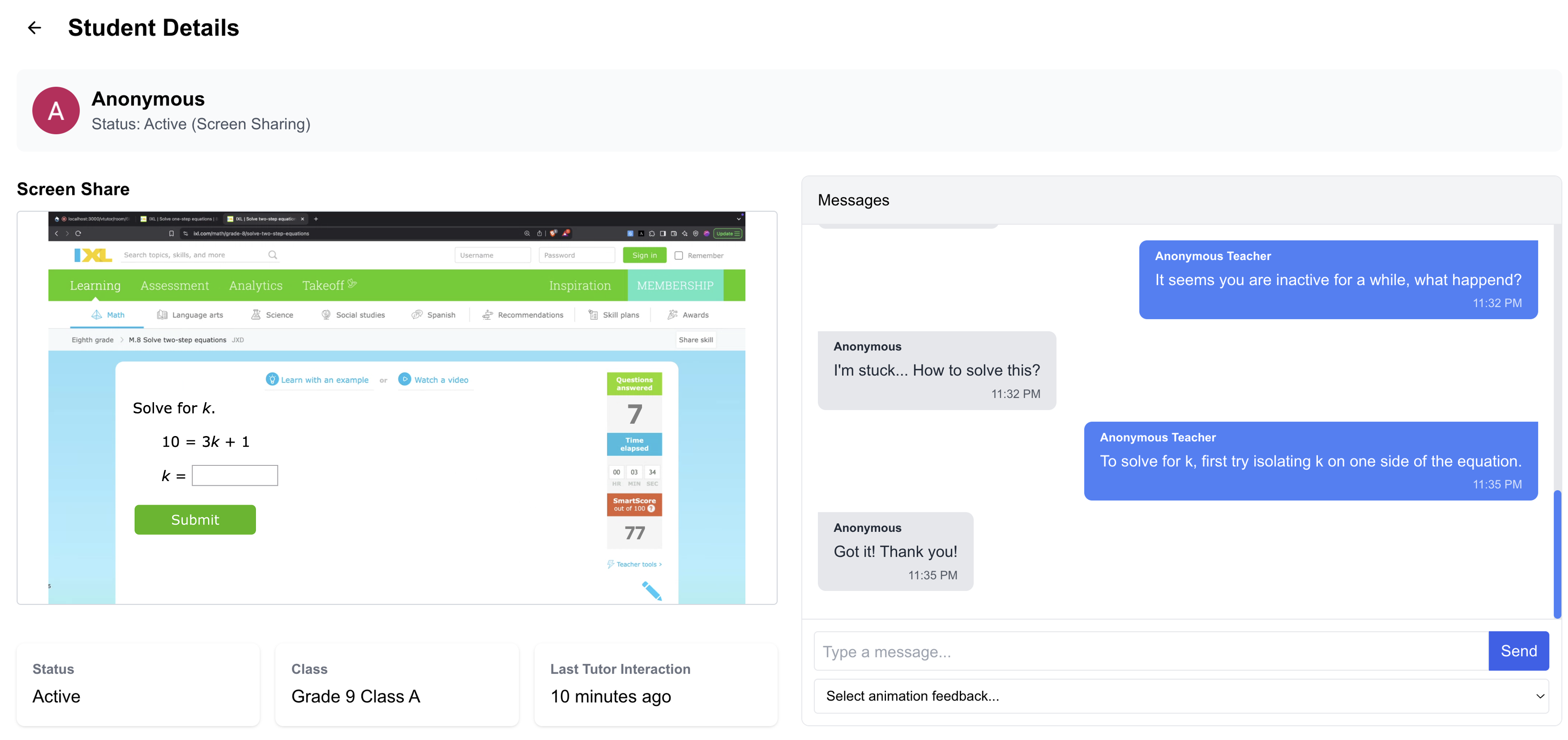}
    \caption{\textbf{Tutor Dashboard and Messaging Interface for individual student.}
    The tutor's view shows an ``Anonymous'' student currently screen-sharing
    IXL, where the student attempts a math problem. On the right, the tutor 
    and student exchange messages in real-time; any tutor messages sent here 
    are spoken aloud by the VTutor avatar on the student's screen. The lower
    panels provide status information (e.g., last tutor interaction), 
    letting the tutor quickly detect off-task behavior and intervene 
    with targeted guidance.}
    \label{fig:tutor_view}
\end{figure}

\subsection{Real-Time Monitoring and Alerts}
While the tutor monitors the \emph{low-resolution} feed of each student, as shown in top right of the \autoref{fig:architecture}, the system logs actions such as mouse clicks, typed answers, and inactivity. If a student has not interacted with their learning application for a configurable duration (e.g., 120 seconds), the thumbnail is flagged in the tutor’s interface. Likewise, repeated incorrect answers can trigger an alert by listening to events emitted from the tutoring system, prompting the tutor to check on that student more closely.

\subsection{Providing Intervention via VTutor Avatar}
When a tutor notices off-task behavior (or receives an automated alert), they can click the relevant thumbnail to:
\begin{enumerate}
    \item \textbf{Enlarge the student’s feed} to observe details of their current task (\autoref{fig:tutor_view}).
    \item \textbf{Type a custom message} or select from a set of predefined hints (e.g., ``Let’s break this down step by step.'').
    \item \textbf{Dispatch the message}, which is spoken aloud by the student’s avatar and optionally displayed as a text bubble.
\end{enumerate}
The avatar engine synchronizes lip movements with text-to-speech, and a short waving animation may play to catch the student’s eye, as shown in \autoref{fig:student_screenshot}. This design keeps the student’s focus on the content while ensuring help is contextual and timely, following elements similar to past interventions addressing disengagement and gaming-the-system behavior in tutoring systems \cite{baker2006adapting}.

\subsection{Ongoing Engagement and Session Closure}
The tutor continuously scans the thumbnails, intervening as necessary. When the session ends, the tutor may disable streaming on the dashboard; students can simply close their browser tabs to terminate the session. All logs and conversation histories are saved in the PostgreSQL database for subsequent review or analytics.

\subsection{Example Scenario.} 
As shown in the \autoref{fig:tutor_view} and \autoref{fig:student_screenshot}, during a middle-school algebra session using IXL, a student repeatedly inputs incorrect attempts for a two-step equation. The tutor’s notice this by viewing student’s thumbnail. Upon clicking it, the tutor sees the student is stuck on ``$10 = 3k + 1$.'' The tutor quickly sends a targeted hint: ``To solve for k, first try isolating k on one side of the equation.'' On the student’s screen, the panda avatar waves and delivers this hint in a friendly, animated voice. The student then re-engages with the problem, typing in the next step of the solution.

\section{Discussion and Future Directions}
\label{sec:conclusion}
\noindent

Hybrid tutoring becomes an increasingly common application area in the field to deliver high-impact tutoring at scale \cite{lin2023personalized,wang2024tutor}. However, past learning analytics and dashboard applications are limited in guiding teacher or tutor attention to directly communicate and attend to multiple students needing help at the same time. Further, past video conferencing software typically requires switching between student screens (e.g., through breakout rooms) which increases tutor's cognitive load \cite{plumlee2006zooming,juola2004task}.

VTutor bridges the gap between personalized tutoring and scalable classroom instruction by combining real-time peer-to-peer screen monitoring with AI-driven avatar interventions. The system empowers educators to oversee multiple learners simultaneously while delivering timely, context-aware support through an engaging virtual agent. By integrating principles from intelligent tutoring systems, teacher dashboards, and embodied conversational agents, VTutor demonstrates how lightweight, browser-based tools can preserve the benefits of one-on-one guidance in large-scale settings.

Challenges remain, including bandwidth optimization and support for low-resource environments, refining avatar expressiveness to better convey empathy (e.g., through personalizing the panda avatar based on student interest; \cite{bernacki2018role}), and enhancing detection algorithms to identify nuanced learner states (e.g., detecting unproductive learning behavior; \cite{baker2006adapting,han2024improving}). One promising application of VTutor is to use generative AI and computer vision models to interpret learner behavior from screen sharing captures \cite{chen2022focusplusdetectlearners,chen2022preliminary}, which could help guide educator attention when overseeing multiple screens at the same time. Future work will explore tighter integration with domain-specific tutoring systems, adaptive gesture generation for avatars, and longitudinal studies to measure VTutor’s impact on learning outcomes. Ultimately, this hybrid approach—blending human oversight with AI-driven scaffolding and learner behavior interpretation—offers a promising path toward high-impact tutoring at scale.

\bibliographystyle{ACM-Reference-Format}
\bibliography{reference}

\end{document}